\documentclass[aps,twocolumn,showpacs,superscriptaddress]{revtex4}
\usepackage[dvips]{graphics,graphicx}

\begin{document}
\title{Bose-Einstein condensates on tilted lattices: coherent, chaotic and 
subdiffusive dynamics}
\author{Andrey R. Kolovsky}
\affiliation{Kirensky Institute of Physics and Siberian Federal University, 
660036 Krasnoyarsk, Russia}
\author{Edgar A. G\'omez}
\affiliation{Fachbereich Physik, Technische Universit{\"a}t Kaiserslautern, 
D-67653 Kaiserslautern, Germany}
\affiliation{Departamento de F\'isica, Universidad Nacional de Colombia, 
Bogot\'a D.C., Colombia}
\author{Hans J\"urgen Korsch}
\affiliation{Fachbereich Physik, Technische Universit{\"a}t Kaiserslautern, 
D-67653 Kaiserslautern, Germany}
\date{\today}

\begin{abstract}
The dynamics of a (quasi)one-dimensional interacting atomic Bose-Einstein condensate  
in a tilted optical lattice is studied in a discrete mean-field approximation, 
i.e., in terms of the  discrete nonlinear Schr\"odinger equation. If the static 
field is varied the system shows a plethora of dynamical phenomena. 
In the strong field limit we demonstrate the existence of (almost) non-spreading 
states which remain localized on the lattice region populated initially and show 
coherent Bloch oscillations with fractional revivals in the momentum space  
(so called quantum carpets). With decreasing field, the dynamics becomes 
irregular, however, still confined in configuration space.  For even weaker 
fields we find sub-diffusive dynamics with a wave-packet width spreading as 
$t^{1/4}$.
\end{abstract}
\maketitle

Recently much attention has been payed to Bloch oscillations (BO) of a Bose- Einstein 
condensate (BEC) of cold atoms in tilted optical lattices (see Refs.
\cite{Ande98,Mors01,57-61,Kolo04a,Zhen04,Ott04,04bloch_bec,Gust08a,Fatt08,Gust08b,09BOBECa}, 
to cite few of the relevant papers). The main question one addresses here, both experimentally and theoretically, is the effect of atom-atom interactions 
on the otherwise perfectly periodic atomic dynamics. Although all experiments on BEC's BO were done for a localized initial conditions, theoretically a simpler case of uniform initial 
conditions, where atoms are delocalized over whole lattice, is usually analyzed 
\cite{57-61,Kolo04a,Zhen04,09BOBECa}. It was found that depending on the system 
parameters a BEC of interacting atoms may show three different regimes of BO 
\cite{09BOBECa}: (i) exponentially decaying BO where the condensate rapidly decoheres, 
which corresponds to unstable (chaotic) BO in the mean-field approach; (ii) persistent BO 
and (iii) modulated BO, which both correspond to stable dynamics in the mean-field approach. 
Assuming the magnitude of a static field to be the control parameter, these three regimes 
refer to weak, moderate, and strong static fields, respectively.

In the present work we extend previous studies to the case of nonuniform initial conditions, 
realized in a laboratory experiment. In what follows we focus on the mean-field analysis, - 
a comparison with quantum mechanical treatment remains an open and challenging problem, as 
it will be discussed in some more detail in the concluding paragraph of the paper.

In the mean-field approximation, the dynamics of a BEC in a tilted optical 
lattice can be described by the discrete nonlinear Schr\"odinger equation (DNLSE)
\begin{equation}
\label{1}
i\;\hbar\dot{a}_l=dFl a_l-\frac{J}{2}(a_{l+1}+a_{l-1}) +g|a_l|^2a_l 
\end{equation}
where $a_l$ is the BEC complex amplitude in the $l$th potential well. 
The static force $F$ leads to a linear increase of the onsite energy
$dFl$ ($d$ is the lattice period), $J$ measures the tunneling transitions 
between the wells and $g$ characterizes the atomic interaction. 
Equation (\ref{1}) appears as a canonical equation of motion  
$i\,\hbar\dot a_l=\partial H/\partial a^*_l$, 
$i\,\hbar\dot a_l^*=-\partial H/\partial a_l$ 
of a `classical' Hamiltonian function
\begin{equation}
H=\sum_ldFl\,|a_l|^2-\frac{J}{2}\sum_l\left(a^*_{l+1}a_l+c.c.\right)
+\frac{g}{2}\sum_l|a_l|^4\,.
\label{1b}
\end{equation}
The norm $\sum_l |a_l|^2$ is conserved under the evolution and we assume $\sum_l |a_l|^2=1$ 
in the following (which assumes that $g$ is proportional the number of atoms), as well as 
units where \,$\hbar=1$ and $d=1$.

The gauge transformation $a_l(t)\rightarrow \exp(-iFlt) a_l(t)$ converts 
the static term in Eq.~(\ref{1}) into a periodic driving with  the Bloch 
frequency $F$,
\begin{equation}
\label{1a}
i\dot{a}_l=-\frac{J}{2}\left( e^{-iFt} a_{l+1} +e^{+iFt} a_{l-1} \right)
+g|a_l|^2a_l  \;,
\label{timepepdgl}
\end{equation}
and a Fourier transform of the amplitudes $a_l$ yields the Bloch-waves representation
$b_k=\frac{1}{\sqrt{L}} \sum_{l=1}^L \exp(-i\kappa l)\,a_l$,
where $\kappa=2\pi k/L$ is the quasimomentum ($-\pi\le\kappa<\pi$). As follows 
from (\ref{1a}), the amplitudes $b_k$ obey the equation
\begin{equation}
\label{blochwave}
i\dot{b}_k=-J\cos(\kappa-Ft)b_k+\frac{g}{L}\!\!\!\sum_{k_1,k_2,k_3} 
\!\!\! b_{k_1}b_{k_2}b^*_{k_3} 
\tilde{\delta}(k_1+k_2-k_3-k)
\end{equation}
where $\tilde{\delta}(k)$ is the Kronecker function modulo $L$.
It is easy to see that the displayed equations of motion can be solved in 
closed form in the case of uniform initial conditions, $a_l(0)=const$, 
yielding  $b_0(t)\sim \exp\big(i\frac{J}{F}\,\sin(Ft)-i\frac{g}{L}t\big)$ and 
$b_{k\ne0}(t)\equiv 0$\,, i.e. ordinary BO.
The stability of this solution with respect 
to weak perturbations has been analyzed in Ref.~\cite{Kolo04a,Zhen04}, 
also in comparison with related phenomena in the many-particle Bose-Hubbard 
model for a small number of particles and lattice sites \cite{09BOBECa}.
\begin{figure}[b!]
\center
\includegraphics[width=8.5cm, clip]{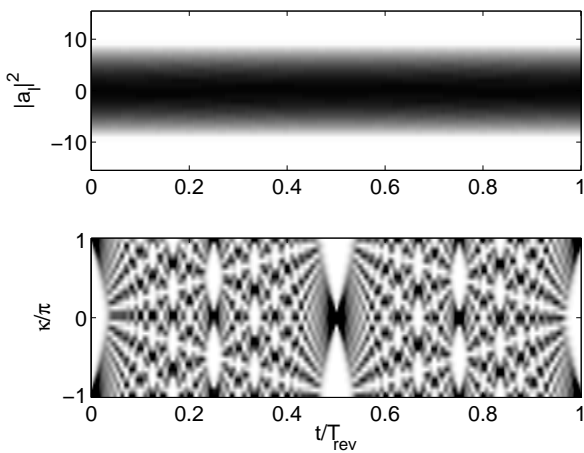}
\caption{Lattice site populations $P_l(t)=|a_l(t)|^2$ (upper panel) and the 
quasimomentum distribution (lower panel) as a function of time in unites of 
the revival time  $T_{\rm rev}=2\pi/g\alpha$ for a force $F=100$ 
(parameters $g=10$, $J=1$). The site populations are frozen and the quasimomentum 
distribution show a highly organized quantum carpet structure.}
\label{plot001}
\end{figure}

Motivated by recent experiments \cite{Gust08b}, 
where a BEC of Cesium atoms was prepared in a harmonic trapping potential 
in the Thomas-Fermi regime, we chose a Thomas-Fermi distribution
\begin{equation}
a_l(0)=\sqrt{\beta-\alpha l^2}
\label{thomasfermi}
\end{equation}
for $\alpha l^2<\beta$ and zero otherwise. We shall use
$\alpha=0.001$, which populates lattice sites between $l=-9$ and $l=+9$ 
(the value of $\beta$ is fixed by normalization). 
The results of a numerical solution of the nonlinear coupled equations 
(\ref{1a}) with initial conditions (\ref{thomasfermi}) and parameters $J=1$, 
$g=10$ are presented in  Figs.~\ref{plot001} and \ref{plot002} as a function of 
time measured in units of $T_{\rm rev}$ as defined in Eq.~(\ref{revtime}) below.  
The site populations 
$P_l(t)=|a_l(t)|^2$ are shown in the upper panel and the quasimomentum 
distributions $|b_k|^2$ in the lower panels. When the force $F$ is varied, 
one observers a characteristic transition from a highly coherent evolution 
for strong fields to an irregular chaotic motion for intermediate fields and
a nonlinear diffusion for weak fields. In the following sections we will discuss
the dynamics in the three different regimes in some detail.

\begin{figure}
\center
\includegraphics[width=8.5cm, clip]{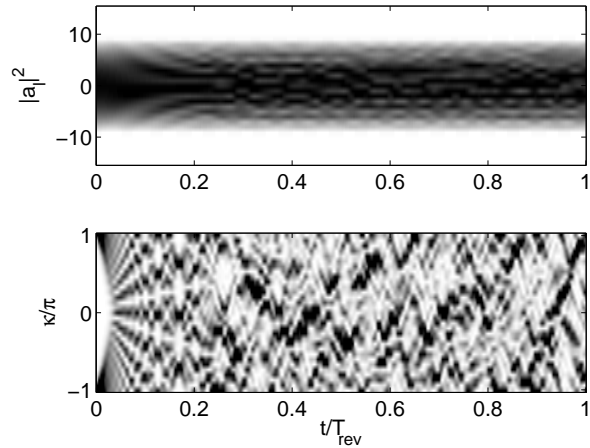}
\caption{Same as Fig.~\ref{plot001}, however for a weaker force $F=10$. The site 
populations are still localized but show fluctuations. The quasimomentum distribution appears 
to be irregular after a transient time.}
\label{plot002}
\end{figure}

\vspace*{1mm}
{\it Coherent evolution.}
For strong fields, as in the case $F=100$ shown in Fig.~\ref{plot001}, the lattice population 
$P_l(t)=|a_l(t)|^2$ is almost frozen without observable structure. On the contrary, the
quasimomentum distribution shows a highly organized pattern, also denoted as
a quantum carpet. The initial distribution is reconstructed at time $T_{\rm rev}$, the revival time, and at rational ratios of $t/T_{\rm rev}$ we find fractional revivals. 
Such highly organized patterns have been observed recently
experimentally for BECs in tilted optical lattices \cite{Gust08b}. 

The origin of these quantum carpets lies in the initial distribution (\ref{thomasfermi}). 
Indeed, in the limit $F\rightarrow\infty$ the site populations are constant 
\cite{09BOBECa}, $|a_l(t)|\approx |a_l(0)|$, and the phases evolve linearly in time:
\begin{equation}
\label{frozen}
a_l(t)\approx a_l(0)\exp\left[-ig|a_l(0)|^2\,t\right] 
\label{amplitudes}
\end{equation}
(see also \cite{04bloch_bec}). In the Bloch-wave representation we find
\begin{equation}
b_k \sim \exp(-ig\beta t)\sum_{l} \exp\big[i(\kappa l +g\alpha l^2t)\big] , 
\quad k=0,\pm 1,\ldots \,,
\label{blochwaveQC}
\end{equation}
an expression already given in \cite{Gust08b}. The most prominent feature 
of such Gaussian sums, which have been analyzed in much detail during recent years
(see, e.g., \cite{Berr96a,Berr96,Gros97b,Stif97b,Berr99}) are the revivals 
and fractional revivals. From Eq.~(\ref{blochwaveQC}) the 
revival time is given by
\begin{equation}
T_{\rm rev}=2\pi/g\alpha
\label{revtime}
\end{equation}
in agreement with the numerical results shown in Fig.~\ref{plot001} for $F=100$.
In comparison with the Bloch period  $T_{\rm B}=2\pi/F$ we have $T_{\rm rev}/T_{\rm B}
=F/g\alpha=10^4$, i.e.~the revival time is much larger than the Bloch period. 
For rational fractions $t/T_{\rm rev}=m/n$ ($n,m$ integer), there appear
$n$ (approximate) copies of the initial distribution of reduced size.
For irrational ratios we find fractal distributions \cite{Berr96}.

It should, however, be pointed out that the limiting strong field behavior in
Eqs.~(\ref{amplitudes}) and (\ref{blochwaveQC}) describes the evolution for finite
values of $F$ only approximately and the coherent evolution shown in Fig.~\ref{plot001} 
is a transient phenomenon for intermediate forces. For $F=10$, as 
shown in  Fig.~\ref{plot002}, the quantum carpets gradually disappear 
after a time interval $T_{\rm coh}\approx 0.2\,T_{\rm rev}$ for $F=10$.
This coherence time was found to vary approximately linear with $F$. 

\vspace*{1mm}
{\it Chaotic evolution.}
The distortion of the quantum carpets in the quasimomentum distribution 
for $t>T_{\rm coh}$ is accompanied by fluctuations in the site populations $P_l(t)$. 
However, lattice populations are still entirely localized on the initial interval and
fluctuate around the initial population $P_l(0)$.  A measure for these fluctuations is 
given by the quantity
$C(t)=\sum_l \big|P_{l+1}(t)-P_l(t)\big|$,
which saturates in time on a level, which increases strongly with decreasing $F$. 
This behavior of $C(t)$ may serves as an indicator the transition from regular to 
irregular dynamics.

An alternative indicator of this transition is the finite time Lyapunov exponent 
\begin{equation}
\lambda(t)=\ln|\delta {\bf a}(t)|/t\;,
\end{equation}
where $\delta {\bf a}(t)$ evolves in the tangent space according to the 
linear equation
\begin{equation}
i\frac{d \delta {\bf a}}{dt}={\cal M}[{\bf a}(t)] \,\delta {\bf a}(t) \,,
\end{equation}
where ${\cal M}$ is the Jacobian matrix of the evolution equation (\ref{timepepdgl}).
We have found that for moderate and weak fields, the finite time Lyapunov exponent 
converges to a positive value after approximately ten Bloch periods, which indicates a chaotic 
motion. For strong fields $F\ge10$ the finite time Lyapunov exponent behaves as $1/t$ during 
the whole simulation time period \cite{remark2}.
%
\begin{figure}[t!]
\center
\includegraphics[width=8.5cm, clip]{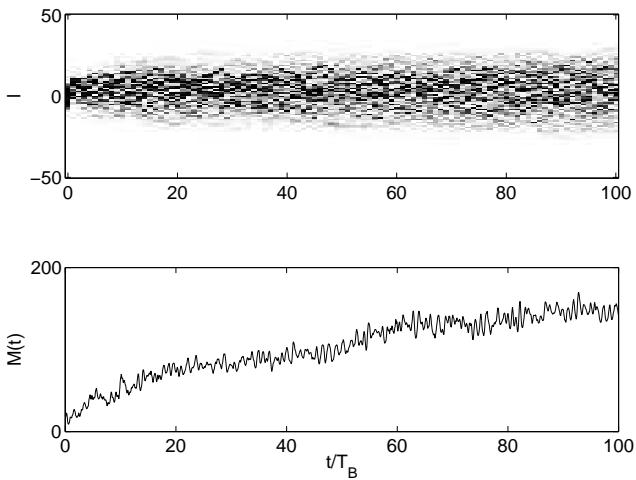}
\caption{Site populations $P_l(t)$ (upper panel) and  dispersion  
$M(t)$ (lower panel) as a function of time in units of the 
Bloch period $T_{\rm B}=2\pi/F$ for $F=0.25$.  The other parameters 
are the same as in Figs.~\ref{plot001}, \ref{plot002} except the upper 
limit of the time axis, which is four times larger.}
\label{my3}
\end{figure}

\begin{figure}
\center
\includegraphics[width=8.5cm, clip]{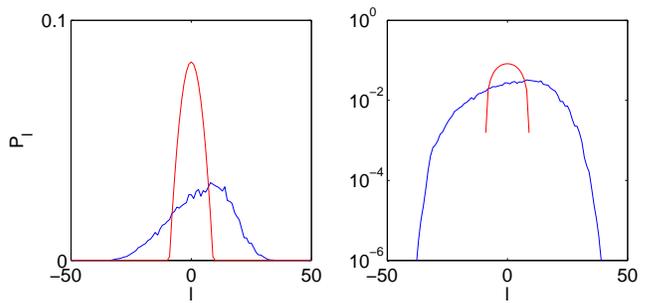}
\caption{Distributions of site populations $P_l$ at the end of the numerical 
simulation discussed in Fig.~\ref{my3} (blue solid line)  compared to the initial distribution 
(red solid line) on linear (left panel) and logarithmic (right panel) scales.}
\label{my4}
\end{figure}

\vspace*{1mm}
{\it Nonlinear diffusion.}
As mentioned above, for a weak field the dynamics of the system is chaotic with no sign of 
the quantum carpet in the quasimomentum distribution and erratic evolution of the site 
populations. This regime is depicted in the upper panel of Fig.~\ref{my3}, where one notices an additional effect not present in Figs.~1, 2 -- the wave packet spreading. More quantitatively this can be seen in the lower panel, which displays the dispersion $M(t)=\sum_l l^2 P_l-(\sum_l l P_l)^2$ of the population distribution. A closer analysis of the functional dependence of $M(t)$ reveals a $\sqrt t$-law with the prefactor increasing with decrease of the static field magnitude. In addition to Fig.~\ref{my3} and for purposes of future reference, Fig.~\ref{my4} shows the distribution of the site populations $P_l$ at the end of the numerical simulations where, to reduce fluctuations of $P_l(t)$, they are averaged over the last 25 Bloch periods. 
\begin{figure}
\center
\includegraphics[width=8.5cm, clip]{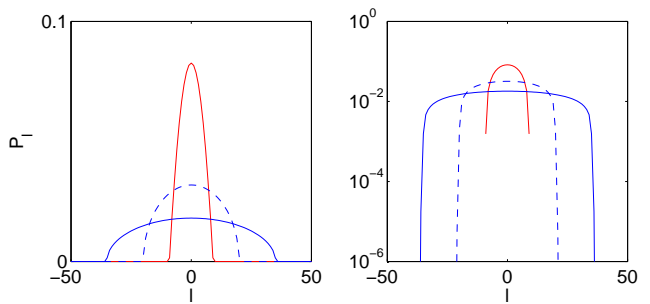}
\caption{Solution of the nonlinear diffusion equation for $\widetilde{D}=50$ on 
linear (left panel) and logarithmic (right panel) scales. Distributions at $t=0$ 
(red solid lines), $t/2\pi=100$ (blue dashed lines), and $t/2\pi=1000$ 
(blue solid lines) are shown. }
\label{my2}
\end{figure}

A qualitative explanation for the observed sub-diffusive spreading is as 
follows. In view of the apparently random behavior of $P_l(t)$ we replace 
in Eq.~(\ref{timepepdgl})  the squared amplitudes $|a_l(t)|^2$ in the 
interaction term by a random variable, 
$|a_l(t)|^2\sim \xi(t)$, with an exponentially decaying correlation function
\begin{displaymath}
R(t-t')=\overline{\xi(t)\xi(t')}=\overline{\xi^2} \exp(-|t-t'|/ \tau) \;.
\end{displaymath}
Assuming the quantity $\overline{\xi^2} \sim \overline{|a_l(t)|^4}$ in the last 
equation to be independent of $l$, this leads to a diffusive spreading 
of the distribution according to the discrete diffusion equation \cite{56}
\begin{equation}
\label{diff1}
\dot{P}_l=D\,(P_{l+1}-2P_l+P_{l-1})
\end{equation}
where the diffusion coefficient $D$ is given by 
\begin{equation}
D=\gamma/({F^2+\gamma^2)\approx \gamma/F^2} \quad  \textrm{with} \quad \gamma=
\overline{\xi^2}\tau \;.
\end{equation}
(Note that Eq.~(\ref{diff1}) is a discretization of the 
continuous diffusion equation $\partial P/\partial t=D\partial^2 P/\partial x^2$ 
and conserves the norm $\sum_l P_l=1$.) Since in the present case the 
quantities  $\overline{|a_l(t)|^4}=P_l^2(t)$ depend on $l$, it is plausible to 
assume that the diffusion coefficient locally depends on $l$ as $D(l)=\tilde{D}P_l^2(t)$. 
Then the site populations $P_l$ obey a nonlinear diffusion equation
\begin{equation}
\label{ndiff1}
\dot{P}_l=\widetilde{D}\,\big(P_{l+1}^3 - 2P_{l}^3 + P_{l-1}^3\big) \;.
\end{equation}
This equation can be considered as a discretization of the continuous nonlinear diffusion equation
\begin{equation}
\label{ndiff2}
\frac{\partial P}{\partial t} = \widetilde{D} \frac{\partial \ }{\partial x} P^\nu 
\frac{\partial P}{\partial x} \;, 
\quad \textrm{for}\quad \nu=2 \;,
\end{equation}
which appears, for example, in the problem of gas diffusion in a porous media \cite{Zeld50}. For the considered case $\nu=2$ one of the exact solutions of 
Eq.~(\ref{ndiff2}) is a semicircular distribution with a radius growing as $t^{1/4}$, 
which implies that the second momentum increases as $M(t)\sim\sqrt{t}$.

The discrete nonlinear diffusion equation (\ref{ndiff1}) seems to inherit  
properties of its continuous counterpart, although we are not aware of any 
formal proof of this statement. As an example, Fig.~\ref{my2} shows the result 
of numerical solution of Eq.~(\ref{ndiff1}) for the initial Thomas-Fermi 
distribution.  Note, that asymptotically the solution is insensitive to the 
particular shape of the initial distribution and one gets the same result, for 
example, for a Gaussian of the same width.

A comparison between Fig.~\ref{my4} and Fig.~\ref{my2} shows that there are 
deviations of the actual profile for the site populations from that predicted by 
the nonlinear diffusion model (\ref{ndiff1}). Nevertheless, the model (\ref{ndiff1}) 
is capable to capture some essential features of the system dynamics, in particular, 
the $t^{1/4}$-law for the wave-packet spreading.

In conclusion, we have considered Bloch dynamics of a BEC in tilted optical lattices 
for the Thomas-Fermi initial profile, a typical initial condition in a 
laboratory experiment. Similar to the case of uniform initial conditions studied 
earlier, the system dynamics is found to strongly depend on the magnitude of a static 
field: it is regular in the strong field limit and chaotic in the weak field limit. 
In the former case of regular dynamics the BEC wave packet is frozen in the 
configuration space and shows quantum carpet evolution in the momentum space, 
a regime already observed in the experiment \cite{Gust08b}. In the latter case of 
chaotic dynamics the BEC spreads sub-diffusively, with the second momentum growing 
approximately as $\sqrt{t}$. This latter regime deserves further studies because of 
its relation to the problem of quantum chaotic diffusion. Indeed, it has been known 
since the pioneering work \cite{Casa78} that quantum interference effects strongly 
modify the classical diffusion due to chaotic dynamics (the so-called phenomenon of 
dynamical localization \cite{Stoe99}). Because the mean-field approach used in this 
work can be considered as a pseudo classical limit of the quantum many-body dynamics 
\cite{09BOBECa}, it is very interesting to compare the above mean-field prediction 
with the behavior of the second momentum calculated on the basis of the Bose-Hubbard 
model, as well as that observed in a laboratory experiment.

Subsequent to the preparation of this
paper we learned about recent work \cite{Flach} where the authors 
address the same problem of different dynamical regimes in biased DNLSE, 
although with a different motivation. As initial conditions an extreme case 
of single site population was mostly considered. 
Among other regimes, a regime with no wave-packet spreading and one with 
sub-diffusive spreading, where $M(t)\sim t^{0.38}$, were reported. 
Taking into account that solutions of a nonlinear equation are typically 
sensitive (even asymptotically) to the type of initial conditions, results 
of the both papers may be considered as consistent.

\vspace*{1mm}
\noindent
{\it Acknowledgments}\\
Support from the Deutsche Forschungsgemeinschaft via the Graduiertenkolleg  
`Nichtlineare Optik und Ultrakurzzeitphysik', Volkswagen Foundation, DAAD and 
Colciencias is gratefully acknowledged.


\begin{thebibliography}{10}

\bibitem{Ande98} 
B.~P.~Anderson and M.~A.~Kasevich, Science {\bf 282} (1998) 1686.

\bibitem{Mors01} 
O.~Morsch, J.~H.~M\"uller, M.~Cristiani, D.~Ciampini, and E.~Arimondo,
Phys. Rev. Lett. {\bf 87} (2001) 140402.

\bibitem{57-61} 
A.~R.~Kolovsky, Phys. Rev. Lett. {\bf 90} (2003) 213002;
A.~Buchleitner and A~.R.~Kolovsky, Phys. Rev. Lett. {\bf 91} (2003), 253002.

\bibitem{Kolo04a}
A.~R. Kolovsky,  preprint: cond-mat/0412195  (2004).

\bibitem{Zhen04}
Yi~Zheng, M.~Kostrun, and J.~Javanainen,  Phys. Rev. Lett.  {\bf 93}  (2004) 230401.

\bibitem{Ott04} 
H.~Ott, E.~de~Mirandes, F.~Ferlaino, G.~Roati, G.~Modugno, and M.~Inguscio, 
Phys. Rev. Lett. {\bf 92} (2004) 160601.

\bibitem{04bloch_bec}
D.~Witthaut, M.~Werder, S.~Mossmann, and H.~J. Korsch,  
Phys. Rev. E  {\bf 71}  (2005)   036625.

\bibitem{Gust08a} 
M.~Gustavsson, E.~Haller, M.~J.~Mark, J.~G.~Danzl, G.~Rojas-Kopeinig, and H.-C.~N\"agerl, 
Phys. Rev. Lett. {\bf 100} (2008) 080404.

\bibitem{Fatt08} 
M.~Fattori, C.~D’Errico, G.~Roati, M.~Zaccanti, M. Jona-Lasinio, M.~Modugno, 
M.~Inguscio, and G.~Modugno, 
Phys. Rev. Lett. {\bf 100} (2008) 080405.

\bibitem{Gust08b}
M.~Gustavsson, E.~Haller, M.~J. Mark, J.~G. Danzl, R.~Hart, A.~J. Daley, and H.-C. N\"{a}gerl, 
arXiv 0812.4836 (2008).

\bibitem{09BOBECa}
A.~R. Kolovsky, E.~M. Graefe, and H.~J. Korsch, arXiv 0901.4719 (2009) 

\bibitem{Berr96a}
M.~V. Berry,  J. Phys. A  {\bf 29}  (1996)   6617.

\bibitem{Berr96}
M.~V. Berry and S.~Klein,  J. Mod. Opt.  {\bf 43}  (1996)   2139.

\bibitem{Gros97b}
F.~Grossmann, J.~M. Rost, and W.~P. Schleich,  
J. Phys. A  {\bf 30}  (1997)  L277.

\bibitem{Stif97b}
P.~Stifter, W.~E.~Lamb Jr., and W.~P. Schleich,  
in {\em Frontiers of Quantum Optics and Laser Physics}, edited by Y.~S. Zhu, M.~S. Zubairy, and M.~O. Scully. World Scientific, Singapore, 1997.

\bibitem{Berr99}
M.~V. Berry,  J. Phys. A  {\bf 32}  (1999)   L329.

\bibitem{remark2}
This does not imply that the Lyapunov exponent is strictly zero, i.e., one should not exclude a possibility that the quasi regular dynamics depicted in Fig.2 is also a transient.

\bibitem{56}
A.~R.~Kolovsky, A.~V.~Ponomarev and H.~J.~Korsch,
Phys. Rev. A {\bf 66} (2002) 053405.

\bibitem{Zeld50}
Ya.~B. Zeldovich and A.~S. Kompaneets,  
in {\em Collection in Honor of the Seventieth Birthday of Academician A. F. Ioffe}, page~61. Izdat. Akad. Nauk SSSR, Moscow, 1950.

\bibitem{Casa78}
G.~Casati, B.~V. Chirikov, F.~M. Izrailev, and J.~Ford,  
in {\em Lecture Notes in Physics, Vol.93}, page 334. Springer, Berlin, 1979.

\bibitem{Stoe99}
H.-J. St\"ockmann,  {\em Quantum Chaos}, Cambridge University Press, Cambridge, 1999.

\bibitem{Flach}
D.~O.~Krimer, R.~Khomeriki, and S.~Flach, arXiv 0904.2867  (2009).

\end{thebibliography}

\end{document}